\documentclass[twocolumn, linenumbers, longbibliography, prl, amsmath, amssymb,superscriptaddress]{revtex4-2}
\usepackage{amsfonts}
\usepackage{graphicx}
\usepackage[colorlinks,linkcolor=blue,citecolor=blue,urlcolor=blue]{hyperref}
\usepackage{float}
\usepackage[normalem]{ulem}

\usepackage{lipsum}

\usepackage{comment}
\usepackage{amsmath,amssymb,graphicx}
\usepackage{soul,xcolor}
\usepackage{textcomp}
\usepackage{siunitx}

\begin{document}
\nolinenumbers

\title{Timekeeping precision enhancements at constant power}

\author{K. J. H. Peters} 
\affiliation{Center for Nanophotonics, AMOLF, Science Park 104, 1098 XG Amsterdam, The Netherlands}

\author{B. Braeckeveldt}
\affiliation{Micro- and Nanophotonic Materials Group, Research Institute for Materials Science and Engineering, University of Mons, 20 Place du Parc, Mons B-7000, Belgium}

\author{B. Maes}
\affiliation{Micro- and Nanophotonic Materials Group, Research Institute for Materials Science and Engineering, University of Mons, 20 Place du Parc, Mons B-7000, Belgium}

\author{S. R. K. Rodriguez}  \email{s.rodriguez@amolf.nl}
\affiliation{Center for Nanophotonics, AMOLF, Science Park 104, 1098 XG Amsterdam, The Netherlands}

\begin{abstract}
We demonstrate precision enhancements in a timekeeping device at constant power and regardless of the operation frequency. Our timekeeping device is a laser-driven coupled-cavity system sustaining limit cycles. We quantify the precision of this device via the standard deviation of the limit cycle period, and demonstrate how it changes when varying the cavity length at constant laser power. Through a phase space analysis of the limit cycle fluctuations, we reveal how the proximity of different bifurcations determines the timekeeping precision of our device regardless of the input power and  oscillation frequency. We expect that, as the miniaturization of computer clocks demands greater energy efficiency in the presence of strong fluctuations, our results can pave to way towards maximizing the precision of such clocks. 
\end{abstract}
\date{\today}
\maketitle

%=========================================
%           INTRODUCTION
%=========================================
Keeping track of time is essential for information processing. Consider, for example, the internal clocks of mammals and  digital computers. Without these clocks, the synchronization and ordering of events necessary for life and computation cannot emerge. A clock's precision is limited by thermodynamics~\cite{barato2015, barato2016, Erker17, Monti18, Milburn20, pearson2021, Ziemkiewicz22, Pietzonka22, gopal2024}. There exists a minimum energy required to maintain a clock's precision given a noise strength~\cite{barato2015, barato2016}.  However, most clocks operate above this ultimate limit, where the relation between energy and precision can be subtle. %fundamental physics. Understanding the underlying mechanisms is key to improve the accuracy of timekeeping, especially under energy constraints and strong fluctuations. Recent efforts in stochastic thermodynamics led to important insights in this direction 

%Intuitively, we expect the energy budget and noise strength to ultimately limit the clock’s precision. 
%In the context of timekeeping and beyond, l

Nonlinear systems supporting limit cycles --- isolated closed orbits in phase space~\cite{strogatz1994} --- are ideal models of clocks. Their self-sustained oscillations can provide a time reference~\cite{jenkins2013}. While not essential for timekeeping, limit-cycle oscillators are superior clocks in noisy environments~\cite{Monti18}. These limit-cycle oscillators have inspired many advances in physics and adjacent fields, including recent theories about their origin~\cite{delPino23}. Limit cycles can emerge in circadian~\cite{Troein09, Monti18, Murugan18}, neural~\cite{makarov2001, Buendia20, braeckeveldt2024}, optical~\cite{orozco1984, marconi2020, Deng20, zambon2020, Abad24}, mechanical~\cite{peters2022, Coulais23}, optomechanical~\cite{metzger2008, Bagheri13, Piergentili21, Livska24}, and  electromechanical~\cite{Villanueva13} systems, to name a few examples. In the past decade, interest in limit cycles surged with the advent of time crystals~\cite{wilczek2012c, Monroe17, Smits18, Heugel19, Taminiau21, Kessler21, Oberreiter21, Kongkhambut22, Zaletel23, Carraro24, Daviet25}: time-periodic states emerging through spontaneous breaking of time-translation symmetry. Such a symmetry breaking also occurs at the onset of a limit cycle, where its phase is randomly selected~\cite{Kongkhambut22}. In the context of both time crystals and clocks, understanding how fluctuations affect self-sustained oscillations is one of the main goals~\cite{Heugel23}. This understanding is necessary to create precise and robust clocks, especially when fluctuations are prominent and energy is limited.
% References specific on thermal or excitability need to come later, connected with our work
%~\cite{orozco1984,marino2004,yacomotti2006,metzger2008,chen2012,brunstein2012,zambon2020}

% ~\cite{wilczek2012, Monroe17}

%~\cite{iemini2018,lledo2019,kessler2019,seibold2020,marconi2020,mattes2023}

%\begin{itemize}\begin{footnotesize}
%\item underdamped (classical) pendulum clock can break TUR~\cite{Pietzonka2022}. Demonstration in electronic system~\cite{gopal2024}

%\item how the amount of dissipated energy affects the precision of the clock? Linear relation between the accuracy and the rate of entropy increase~\cite{erker2017} (quantum system)

%\item above point has been confirmed to exist in nanoscale electromechanical clocks~\cite{pearson2021} 

%\item same as above two points, but for mechanical clock: demonstrates that the clock’s precision scales linearly with the rate of its entropy production.~\cite{ziemkiewicz2022}

%\item ``thermodynamics dictates a trade-off between the amount of dissipated heat and the clock’s performance in terms of its accuracy and resolution''~\cite{erker2017}

%\item ``All clocks, classical or quantum, are open non equilibrium irreversible systems subject to the constraints of thermodynamics. Using examples I show that these constraints necessarily limit the performance of clocks and that good clocks require large energy dissipation. \emph{For periodic clocks, operating on a limit cycle, this is a consequence of phase diffusion}.''~\cite{milburn2020}

%\item 
%\end{footnotesize}\end{itemize}

\begin{figure}[!]
    \includegraphics[width=\columnwidth]{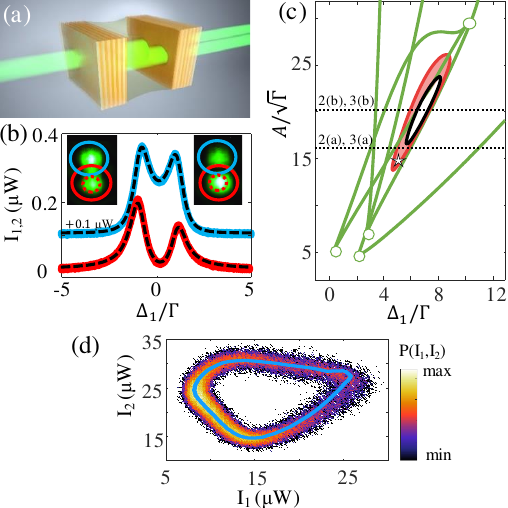}
    \caption{(a) Two oil-filled coupled cavities. (b) Intensities transmitted by the laser-driven ($I_1$, solid red) and undriven ($I_2$, solid blue, displaced for clarity) cavities as function of the laser-cavity detuning $\Delta_1/\Gamma$, in the linear regime. Black dashed curves are steady-state solutions of Eq.~\ref{eq:IDE}. Insets: transmission images of the bonding (left) and antibonding (right) modes. Solid red  and blue circles enclose the driven and undriven cavity, respectively. Dashed red circles indicate the location of the driving laser. (c) Curves of saddle-node (green), Hopf (red) and homoclinic (black) bifurcations, predicted by Eq.~\ref{eq:IDE}. Cusp bifurcations are shown as open green circles. Limit cycles are expected in the red shaded region. (d) Experimentally-obtained joint probability distribution of $I_{1,2}$ at the driving conditions indicated by the star in (c). The blue curve is the experimentally-obtained mean limit cycle.}\label{fig:1}
\end{figure}

%=========================================
%           MAIN
%=========================================
In this Letter we demonstrate timekeeping precision enhancements at zero energy cost. We present experiments and theory for a laser-driven coupled cavity system sustaining limit cycles and thereby functioning as a clock. Contrary to the intuition that a clock's energy consumption and frequency determine its precision, we observe substantial precision enhancements at constant power and with minor frequency changes. We achieve this by adjusting the cavity length, which determines the proximity of the limit cycle to different bifurcations. As those bifurcations influence the limit cycle fluctuations, the timekeeping precision is thereby modified. Our experimental results are supported by numerical simulations, enabling us to analyze fluctuations in phase space. We conclude by offering perspectives for our results.

% End matters
%  Model parameters: $\beta=\Gamma/800$, $\kappa_\mathrm{L}=\Gamma/2$, $U=\Gamma/40$, $\tau=5000\Gamma^{-1}$, $A=0.25\sqrt{\Gamma}$, $J=\Gamma$, $\delta=\omega_2-\omega_1=-0.25\Gamma$, $\rho=0.95$.

%Experiments involving single-mode cavities with a thermo-optical nonlinearity have been successfully modelled by such a non-instantaneous nonlinearity~\cite{geng2020,peters2021}. Here, we model our system in a similar way but with two major  by two such identical cavities linearly coupled with strength $J$~\cite{braeckeveldt2024}.

% a thermo-optical nonlinearity, which is non-instantaneous

Figure~\ref{fig:1}(a) illustrates our experimental system: two coupled side-by-side microcavities (details in Appendix A) filled with olive oil and driven by a continuous wave laser. The oil partly absorbs the laser light, warms up, expands, and its refractive index thereby changes. This makes the cavity optical response nonlinear~\cite{geng2020,peters2021}. Taking this into account, the light fields $\alpha_{1,2}$ in our oil-filled coupled cavities obey the following  equations of motion in a frame rotating at the laser frequency $\omega$:
\begin{widetext}
    \begin{equation}\label{eq:IDE}
            i\dot{\alpha}_j(t) = \left[-\Delta_j-i\left(\frac{\Gamma}{2}+\beta\left|\alpha_j(t)\right|^2\right)+U\int_0^t ds\;K(t-s)\left|\alpha_j(s)\right|^2\right]\alpha_j(t)
            + i\sqrt{\kappa_\mathrm{L}}\left[1-(-1)^j\rho\right]A - J\alpha_{3-j}(t) + D\xi_j(t)
    \end{equation}
\end{widetext}
% Why do we need this staement when little delta is not in the equation?
% The cavity-cavity detuning is $\delta=\omega_2-\omega_1=\Delta_2-\Delta_1$, such that $\Delta_1=\Delta$ and $\Delta_2=\Delta+\delta$. 
$\Delta_j=\omega-\omega_j$ is the frequency detuning between the laser and the $j^{th}$ cavity resonance $\omega_j$. $\Gamma=\gamma+\kappa_\mathrm{L}+\kappa_\mathrm{R}$ is the total loss rate of each cavity, including absorption $\gamma$  and input-output losses through the left (right) mirror $\kappa_\mathrm{L}$ ($\kappa_\mathrm{R}$). $\beta$ is the strength of nonlinear losses, necessary to reproduce our experimental observations (see Supplemental Material~\cite{supp}). $U$ is the thermo-optical nonlinearity strength, and $K(t)=\mathrm{exp}(-t/\tau)/\tau$ a memory kernel characterizing the oil's  thermal relaxation time. $\tau$ is the thermal relaxation time as well as  the memory time of the system. Following the approach in Ref.~\cite{peters2021} and summarized in Appendix B, we estimate $\tau = 1.2~\mu$s for our system. $\gamma$, $\beta$, $U$, and $\tau$ are equal for both cavities since they are both filled with the same oil. $\Gamma$ could in principle be different for each cavity, since small asymmetries could imbalance $\kappa_\mathrm{L,R}$ for the two cavities. However, we observe no signatures of this effect in our experiments. $A$ is the laser amplitude, with $\rho\in\left[-1,1\right]$ the driving imbalance between cavities. $J$ is the intercavity coupling rate. $D\xi_j(t)=D\left[\xi_j^R(t) +i\xi_j^I(t)\right]/\sqrt{2}$ are Gaussian white noises with zero mean, correlations $\left\langle \xi_i(t)\xi_j(t')\right\rangle=\delta_{ij}\delta(t-t')$ and variance $D^2$, accounting for amplitude and phase noise in the driving laser. The same stochastic terms account for fluctuations in the intracavity field, assuming all noise sources are white, Gaussian, and additive.  Details about our model and its numerical implementation are reported in Appendix C.

Figure~\ref{fig:1}(b) shows the experimental intensity transmitted by each cavity when driving only the lower cavity at a low laser power, ensuring linear response. We also plot the steady-state solutions of Eq.~\ref{eq:IDE}, with parameter values chosen to fit our experimental data. Fitting the model to our data in this way, we obtained the values of $\rho$, $J$, and the inter-cavity detuning  $\delta = \Delta_{2}- \Delta_{1}$ corresponding to our experiments. Figure~\ref{fig:1}(c) shows the two-parameter bifurcation diagram for our system, obtained via numerical continuation using the Matlab toolbox MatCont~\cite{dhooge2003}. In Supplemental Material we present experimental signatures of  all predicted bifurcations~\cite{supp}. Limit cycles are expected in the shaded red region of Fig.~\ref{fig:1}(c). Measuring the transmitted intensities $I_{1,2}$ in that region, at the location of the star in Fig.~\ref{fig:1}(c), we indeed observe a limit cycle as shown in Fig.~\ref{fig:1}(d). The mean evolution (blue curve) forms an isolated closed orbit in the $I_{1,2}$ plane, and the joint probability distribution  $P(I_1, I_2)$ (encoded in color) forms a ring around it. The width of that ring is determined by the noise variance.

\begin{figure}[!t]
    \includegraphics[width=\columnwidth]{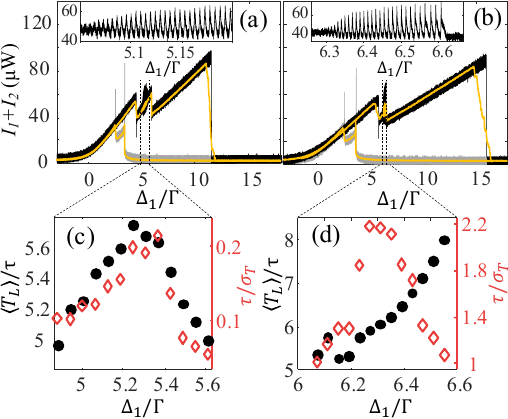}
    \caption{(a) and (b) show the intensity transmitted by both cavities when scanning the detuning forward (black) and back (gray) with a fixed laser power of $7.9$~mW and $9.8$~mW, respectively. Black cuves are a single shot, and yellow curves are the average of 40 scans. Insets: Zoom into the oscillations. (c,d) Average oscillation period (black dots) and precision (red diamonds) as a function of $\Delta_1/\Gamma$ for driving conditions corresponding to (a,b).}\label{fig:2}
\end{figure}

% Which detuning is in Fig. 2a,b

Figures~\ref{fig:2}(a,b) show the total transmitted intensity $I_1+I_2$, while scanning the cavity length in the nonlinear regime, at 7.9 mW and 9.8 mW respectively. These two laser powers are indicated by the dashed lines in Fig.~\ref{fig:1}(c). Black  and gray curves are single scans whereby the detuning increases and decreases, respectively. Yellow curves are the average of $\sim 40$  scans. For both laser powers we observe a large hysteresis, and several jumps due to saddle-node bifurcations corresponding to the green curves in Fig.~\ref{fig:1}(c). We also observe limit cycles within certain detuning ranges [$4.95 \lesssim \Delta_1/\Gamma \lesssim5.6$ in Fig.~\ref{fig:2}(a) and  $6.1 \lesssim \Delta_1/\Gamma \lesssim 6.55$  in Fig.~\ref{fig:2}(b)], as shown for example in the insets of Figs.~\ref{fig:2}(a,b). In Supplemental Material~\cite{supp} we show that, for both laser powers, the oscillation amplitude initially increases as the detuning increases. Upon approaching a Hopf bifurcation in Fig.~\ref{fig:2}(a), the oscillations gradually decrease in amplitude. In contrast,  upon approaching a homoclinic bifurcation in Fig.~\ref{fig:2}(b), the oscillations terminate abruptly and their period diverges. At that homoclinic bifurcation [black curve in Fig.~\ref{fig:1}(c)], the limit cycle collides with a saddle (unstable fixed point) and disappears.

We now focus on the detuning ranges supporting limit cycles, enclosed by dotted lines in Figs.~\ref{fig:2}(a,b). Figures~\ref{fig:2}(c,d)  show the average limit cycle period $\langle T_L \rangle$ and its inverse standard deviation $\sigma_T^{-1}$, both referenced to $\tau$. We use $\tau / \sigma_T$ to quantify the timekeeping precision: the smaller the variance of the limit cycle period, the greater the precision of a timekeeping device. The limit cycles in Fig.~\ref{fig:2}(c) are bound by Hopf bifurcations on both sides. Within this detuning range, $\tau / \sigma_T$  changes by a factor of four while  $\langle T_L \rangle /\tau$ changes by merely $\sim10\%$. Already this result demonstrates substantial timekeeping precision enhancements at constant power and minor frequency changes. Interestingly, the precision is maximized when the operation frequency is minimized. This opposes the widely-held expectation that a clock's precision is proportional to its frequency. Figure~\ref{fig:2}(d), corresponding to a larger laser power, also shows substantial timekeeping precision enhancements at an intermediate detuning.  $\tau / \sigma_T$  follows a trend similar to that in  Fig.~\ref{fig:2}(c),  reaching a maximum at an intermediate detuning. However, unlike in  Fig.~\ref{fig:2}(c), $\langle T_L \rangle /\tau $ increases steadily with detuning.  According to theory, the limit cycle period should diverge upon  approaching the homoclinic bifurcation on the right.  Figure~\ref{fig:2}(d) indeed shows a progressive growth of  $\langle T_L \rangle /\tau $, in qualitative agreement with theory. However, nanoscale vibrations of our experimental cavity setup prevent us from further approaching the homoclinic bifurcation and hence the divergence of the limit cycle period. Overall, Figs,~\ref{fig:2}(c) and ~\ref{fig:2}(d), combined, demonstrate the generality of the phenomenon we study: timekeeping precision variations at constant power. 

\begin{figure}[!t]
    \includegraphics[width=\columnwidth]{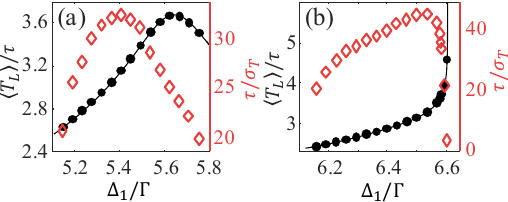}
    \caption{(a) Average oscillation period (black dots) and precision (red diamonds) versus $\Delta_1/\Gamma$ for a constant laser amplitude $A=3.9A_c$. Black curves are obtained via numerical continuation of the limit cycle. (b)  Same as (a) but for $A=4.9A_c$.}\label{fig:3}
\end{figure}

% END MATTERS
%and standard deviation of the noise $D=0.1\sqrt{\Gamma}$

Next we elucidate the physics of our system using our model. To this end, we first validated our model (see Supplemental Material~\cite{supp}) by reproducing our experimental observations across a range of detunings and laser powers. Figures~\ref{fig:3}(a,b) present an analysis similar to that in Figs.~\ref{fig:2}(c,d), but for numerical data. Here, the laser amplitude $A$ is referenced to the critical value for bistability, $A_c$ (see Appendix C). We calculated  $\langle T_L \rangle /\tau $ via numerical continuation, and $\tau / \sigma_T$ via stochastic simulations. For both laser powers we reproduce our three main observations: i)  $\tau / \sigma_T$ is maximized at an intermediate detuning; ii) $\langle T_L \rangle /\tau $ slightly decreases upon approaching a Hopf bifurcation [Fig.~\ref{fig:3}(a)], and increases more substantially upon approaching a homoclinic bifurcation [Fig.~\ref{fig:3}(b)]; iii)   $\tau / \sigma_T$ does not, in general, correlate with  $\langle T_L \rangle /\tau $. We also notice three discrepancies between experiments and simulations: i) The values of $\tau/\sigma_T$ are different; ii) peaks in $\langle T_L \rangle /\tau$ and $\tau / \sigma_T$ are slightly shifted in detuning; iii) $\tau / \sigma_T$ drops substantially upon approaching the homoclinic bifurcation in Fig.~\ref{fig:3}(b), where the limit cycle period diverges.  We believe that discrepancies i) and ii) can be attributed to our use of a much shorter thermal relaxation time in simulations ($\Gamma\tau=5000$) compared to experiments ($\Gamma\tau \sim 10^5$). The shorter relaxation time was necessary due to computer memory limitations, as explained in Appendix C. Discrepancy iii), we believe, is due to the intrinsic experimental difficulty in approaching a homoclinic bifurcation without crossing it. The three discrepancies are nonetheless inconsequential to our study, focused on understanding qualitative changes in timekeeping precision at constant input power. Those changes are similarly evident in both experiments and simulations.

%The precision $\tau/\sigma_T$, obtained from stochastic simulations, also aligns qualitatively with our experimental findings. \KP{We expect deviations to be due to} the much shorter thermal relaxation time used in simulations ($\Gamma\tau=5000$) compared to experiments ($\Gamma\tau \sim 10^5$), as the thermal relaxation time heavily affects limit cycles and their bifurcations~\cite{peters2022,braeckeveldt2024}. However, simulations using the experimental thermal relaxation time are not feasible due to computer memory and time constraints.

\begin{figure}[!t]
    \includegraphics[width=\columnwidth]{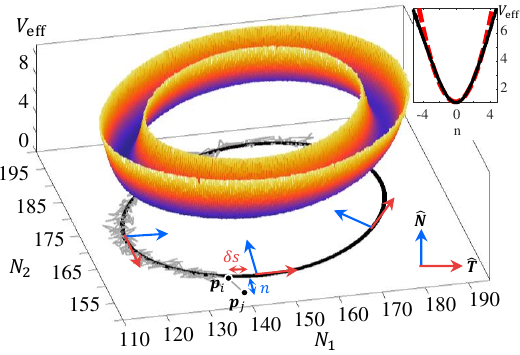}
    \caption{Colored surface: effective potential $V_\mathrm{eff}$  for the limit cycle. Black curve: deterministic limit cycle. Gray curve: part of one stochastic limit cycle. Red and blue arrows: tangent and normal unit vectors, $\mathbf{\hat{T}}$ and $\mathbf{\hat{N}}$. Black dots labeled $\mathbf{p}_{i,j}$ indicate the last two points of the trajectory, illustrating  the definitions of normal deviation $n$ and arclength change $\delta s$.     Inset: $V_\mathrm{eff}$ versus $n$ in black. Red dashed curve is a parabolic fit.}\label{fig:4}
\end{figure}

% PARAMETER VALUES IN END MATTERS
% obtained from $8000$ stochastic orbits for $A=3.9A_\mathrm{c}$, $\Delta_1=5.54\Gamma$ and $D=0.1\sqrt{\Gamma}$

We now explain our observations, which are related to the enhancement of fluctuations near a bifurcation~\cite{dykman1980,knobloch1983,meunier1988}. To this end, we analyze intensity fluctuations in the directions normal and tangent to the deterministic limit cycle in the plane of transmitted intensities, as illustrated in Fig.~\ref{fig:4}. We performed this analysis on numerical data because our experimental setup does not provide a sufficiently smooth limit cycle resembling the deterministic one, even after substantial averaging. The black curve in Fig.~\ref{fig:4} is the deterministic limit cycle. The gray curve is a typical stochastic limit cycle, with its last two points labeled $\mathbf{p}_{i,j}$. We set up a moving reference frame with unit vectors $\mathbf{\hat{T}}$ and $\mathbf{\hat{N}}$ in the directions tangent and normal to the deterministic limit cycle, respectively. Fluctuations normal to the deterministic limit cycle are defined as the distance $n$ from a point $\mathbf{p}_j$ on the stochastic trajectory to the deterministic limit cycle. $n$ therefore corresponds to the projection of $\mathbf{p}_j$ onto $\mathbf{\hat{N}}$ (indicated for the last point on the stochastic trajectory in Fig.~\ref{fig:4}).

Based on 8000 stochastic limit cycles, we obtained a joint probably distribution $P(N_1, N_2)$.  $N_1=|\alpha_1|^2$ and $N_2=|\alpha_2|^2$ correspond to the intensities $I_1$ and $I_2$, respectively.  We can then define an effective potential $V_\mathrm{eff} = -\ln[P(N_1, N_2)]$ for the limit cycle, shown as a colored surface in Fig.~\ref{fig:4}. The local curvature in the normal direction is computed as $k = \frac{d^2 V_\mathrm{eff}}{dn^2}\big|_{n=0}$. Intuitively, $k$ can be interpreted as the spring constant of the restoring force in the direction normal to the mean limit cycle. Thus, the stochastic limit cycle is more strongly confined when $k$ is larger.  Assuming a fixed noise variance, deviations in the normal direction are less prominent under strong confinement of the limit cycle. 

\begin{figure}[!t]
    \includegraphics[width=\columnwidth]{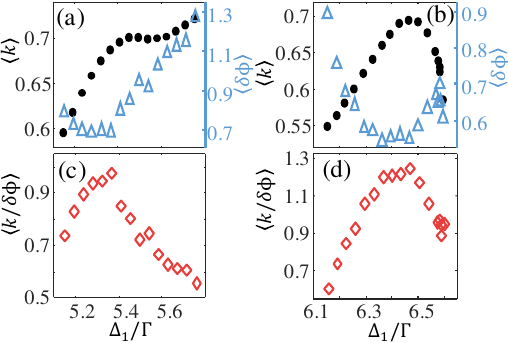}
    \caption{(a,b) Mean curvature $\left\langle k \right\rangle$ (black circles) and phase shift $\left\langle \delta\phi \right\rangle$ (blue triangles) versus $\Delta_1/\Gamma$   at constant power. All model parameters for (a) and (b) are as in Fig.~\ref{fig:3}(c) and ~\ref{fig:3}(d), respectively.   The ratio $\left\langle k/\delta\phi \right\rangle$, which qualitatively reproduced the experimentally-observed timekeeping precision, is plotted in (c) and (d) for the driving conditions in  (a) and (b), respectively.} 
\label{fig:5}
\end{figure}

Figure~\ref{fig:5}(a,b) show the curvature averaged over the full cycle, $\left\langle k\right\rangle$, as a function of $\Delta_1/\Gamma$ for the two laser powers under consideration. In Fig.~\ref{fig:5}(b) we observe how $\left\langle k\right\rangle$ closely follows the trend of $\tau / \sigma_T$ in Fig.~\ref{fig:3}(b). However, $\left\langle k\right\rangle$ in Fig.~\ref{fig:5}(a) shows qualitatively different behavior from $\tau / \sigma_T$ in Fig.~\ref{fig:3}(a). Notice how $\left\langle k\right\rangle$ increases with $\Delta_1/\Gamma$. This trend reproduces the initial rise in $\tau / \sigma_T$  in Fig.~\ref{fig:3}(a), but does not explain the decrease at large detuning. Clearly,  $\left\langle k\right\rangle$ cannot explain the dependence of the timekeeping precision on detuning for this laser power. A more general explanation, capturing the behavior at both laser powers, is needed.    

While $\left\langle k\right\rangle$ influences fluctuations normal to the deterministic limit cycle, fluctuations in the tangent direction can lead to phase diffusion. Stochastic limit cycles can thus obtain a phase difference relative to the deterministic limit cycle, affecting the timekeeping precision. To compute this phase difference we use the arclength change $\delta s$ (see Fig.~\ref{fig:4}). $\delta s$ is defined as the displacement $\mathbf{p}_j - \mathbf{p}_i$ projected onto the tangent unit vector $\mathbf{\hat{T}}$. The accumulated phase difference during one cycle is then $\delta\phi = 2\pi s/\ell$, with $s$ the total arclength of one stochastic cycle ($\delta s$ integrated over the full orbit) and $\ell$ the total arclength of the deterministic limit cycle. Phase diffusion increases with $\delta\phi$, in turn decreasing the timekeeping precision.

Figures~\ref{fig:5}(a,b) show the average accumulated phase difference $\left\langle \delta\phi\right\rangle$. For both laser amplitudes, $\left\langle \delta\phi\right\rangle$ first decreases and then increases with $\Delta_1/\Gamma$. The rise in $\left\langle \delta\phi\right\rangle$ explains the drop in $\tau / \sigma_T$  at large $\Delta_1/\Gamma$. However, in Fig.~\ref{fig:5}(b), the magnitude of the increase in $\left\langle \delta\phi\right\rangle$ at large $\Delta_1/\Gamma$ is much smaller than that of the drop in $\tau / \sigma_T$  observed in Fig.~\ref{fig:3}(b). This suggests that both the potential's curvature and phase diffusion influence the timekeeping precision. Indeed, Figs.~\ref{fig:5}(c,d) show that the ratio $\left\langle k/\delta\phi\right\rangle$ follows the same trends observed for $\tau / \sigma_T$  in Fig.~\ref{fig:3}(a,b). This reproduction of our experimental observations, for both laser powers, suggests that both tangent and normal fluctuations are relevant.  In our system, the strength of both types of fluctuations depends on the laser-cavity detuning, which in turn determines the proximity of bifurcations. While our explanation is only qualitative, we highlight that we considered several other hypotheses to explain our results. However, none of these hypotheses yielded positive results. These hypotheses are related to the intracavity intensity in the limit cycle regime, as well as the oscillation amplitude and phase space velocity of the limit cycle. A detailed discussion of those alternative (unsuccessful) hypotheses and accompanying analysis of our experimental data are presented  in Supplemental Material~\cite{supp}. 

%The small deviation between the detuning of the maxima in $\tau/\sigma_T$ and $\left\langle k/\delta\phi\right\rangle$ could be due to changes in the coupling between $n$ and $\delta s$ with detuning~\cite{sheth2018}.

In summary, we demonstrated how the precision of a timekeeping device can increase at constant input power and with minor frequency changes. We achieved this by controlling the proximity of our device to different bifurcations, where fluctuations degrade the timekeeping precision. To conclude, we offer two perspectives. One is to use our coupled-cavity system to test fundamental results in statistical physics, such as thermodynamic uncertainty relations constraining the precision of clocks~\cite{marsland2019}. This is an area with many theoretical results~\cite{barato2015, barato2016, Erker17, Monti18, Milburn20, pearson2021, Ziemkiewicz22, Pietzonka22, gopal2024} but few experiments. Our work  demonstrates the level of control required for experiments in that direction. These experiments could, in turn, reveal fundamental limits to the precision of clocks undergoing non-Markovian dynamics, as our system. The second perspective we offer is related to the observation of limit cycles in the vicinity of a homoclinic bifurcation, as shown in Fig.~\ref{fig:2}(b). That regime is ideal for realizing the hallmark behavior of neurons, namely excitability --- the ability to release energy suddently upon a small stimulus~\cite{braeckeveldt2024}. The realization of excitability in coupled-cavity systems would be a first step towards the realization of an all-optical spiking neural network.

%=========================================
%           CONCLUSION
%=========================================

\section*{Acknowledgments}
\noindent This work is part of the research programme of the Netherlands Organisation for Scientific Research (NWO). We thank Anne Spakman for preliminary measurements, Bart Verdonschot for discussions, and Aur\'elien Trichet for providing the structured  mirror used in our experiments. S.R.K.R. acknowledges an ERC Starting Grant with project number 852694.  B.B. and B.M. acknowledge support by FNRS-FRIA.
% \end{acknowledgement}
\bibliography{references.bib}

\section*{End Matter}
\textit{Appendix A: Experimental system---}The cavity mirrors are made of distributed Bragg reflectors (DBRs) on a glass substrate. The DBRs have 99.8\%  reflectance at 532 nm, the laser wavelength. The mirrors are aligned and positioned using piezoelectric actuators. One of the mirrors contains two partially overlapping concave features. These were milled with a focused ion beam on the glass substrate prior to the deposition of the DBR, as described in Ref.~\onlinecite{Trichet15}. Each concave feature makes a plano-concave cavity. The cavity modes couple via their mutual field overlap~\cite{dufferwiel2015, Flatten16}. This results in bonding and anti-bonding resonances, evident in the linear transmission spectrum in Fig.~\ref{fig:1}(b). The transmission through each concave mirror is measured by a separate avalanche photodetector. Details about our setup and measurement procedure are in Supplemental Material.

\begin{figure}[!b]
    \centerline{\includegraphics[width=\columnwidth]{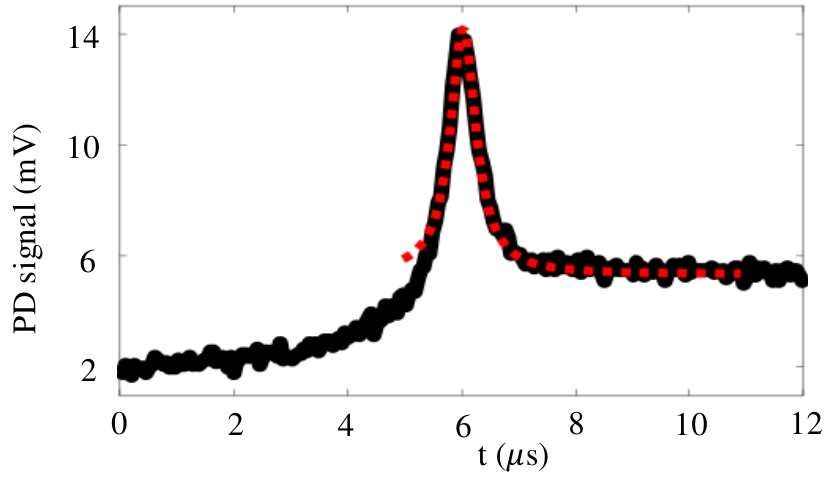}}
    \caption{Single shot measurement of the transmitted intensity of the driven cavity (black) while the input laser is modulated by a chopper. Lorentzian fit to the overshoot is show as a red dotted curve.}\label{fig:Memory}
\end{figure}

\textit{Appendix B: Thermal Relaxation Time---} We determined the thermal relaxation time $\tau$ of our oil-filled cavity system using the procedure introduced in Ref.~\cite{peters2021}. The pump power is modulated in a step-like fashion using a chopper, while keeping the cavity length constant. Figure~\ref{fig:Memory} shows in black a single-shot measurement of the transmitted intensity of the driven cavity as function of time. Fitting a Lorentzian (red dotted curve) to the overshoot, we find a thermal relaxation time of $\tau=1.2\pm 0.1~\mu$s.  This value is smaller than the $16~\mu$s reported in Ref.~\cite{geng2020} for a single-mode oil-filled cavity. We attribute this difference to a smaller mode volume for the experiments reported in this manuscript. The smaller mode volume is due to a tighter confinement in the transverse direction by concave mirrors with a smaller radius-of-curvature, as well as in the longitudinal direction by a shorter cavity length. A tighter confinement is indeed expected to reduce the thermal relaxation time ~\cite{stein2019}.

\textit{Appendix C: Model---}For our stochastic simulations and numerical parameter continuation it is convenient to decompose $\alpha_j=u_j+iv_j$, with $u_j=\mathrm{Re}(\alpha_j)$ and $v_j=\mathrm{Im}(\alpha_j)$, and to define $w_j=U\int_0^t ds\;K(t-s)\left|\alpha_j(s)\right|^2$, such that Eq.~\ref{eq:IDE} becomes
\begin{widetext}
    \begin{equation}\label{eq:ODE}
        \begin{split}
            \dot{u}_j&=-\left(\frac{\Gamma}{2}+\beta N_j\right)u_j-\left[\Delta_j-w_j\right]v_j-Jv_{3-j}+\sqrt{\kappa_\mathrm{L}}\left[1-(-1)^j\rho\right]A+\frac{D}{\sqrt{2}}\xi_j'\\
            \dot{v}_j&=-\left(\frac{\Gamma}{2}+\beta N_j\right)v_j+\left[\Delta_j-w_j\right]u_j+Ju_{3-j}+\sqrt{\kappa_\mathrm{L}}\left[1-(-1)^j\rho\right]A+\frac{D}{\sqrt{2}}\xi_j''\\
            \dot{w}_j&=\left(U N_j-w_j\right)/\tau,
        \end{split}
    \end{equation}
\end{widetext}
with $N_j=\left|\alpha_j\right|^2=u_j^2+v_j^2$. Stochastic simulations are performed using a fourth order Runge-Kutta algorithm with time increments of $\Gamma/10$ and a noise standard deviation of $D=\sqrt{\Gamma}/10$. Steady-state solutions (obeying $\dot{u}_j=\dot{v}_j=\dot{w}_j=0$) and bifurcation diagrams are obtained using numerical parameter continuation using MatCont~\cite{dhooge2003}.

The values of parameters in Eq.~\ref{eq:ODE} were determined as follows. First note that $\Gamma$ is fixed by the resonance linewidth at low pump power. The coupling strength $J$, intercavity detuning $\delta$, and pump imbalance $\rho$ are obtained by fitting the steady-state photon numbers $|\alpha_j|^2$ to the measurements in the linear regime [dashed curves in Fig.~\ref{fig:1}(b)]. From this we obtain $J/\Gamma=1$, $\delta/\Gamma=-0.25$ and $\rho=0.95$. As explained in Ref.~\cite{geng2020}, the value of $A$ only matters relative to the critical value for bistability, $A_c$, while $\kappa_L$ only rescales the pump amplitude. By performing a series of measurements at increasing pump power, we determine the minimum power $A_c^2$ required for bistability, allowing us to reference all measurements to this value. The strength of the nonlinear dissipation $\beta=\Gamma/800$ is obtained by fitting the detuning range at which oscillations occur for $A=3.9A_c$, keeping all other parameters fixed. The measured thermal relaxation time $\tau=1.2~\mu$s is $\sim$6 orders of magnitudes larger than $\Gamma^{-1}$, much too long to be able to simulate given computer memory and time restraints. In our calculations and simulations we therefore take $\tau=5000\Gamma^{-1}$, much longer than $\Gamma^{-1}$, but short enough that it can be simulated within in reasonable time. We note, however, that it is shown in Ref.~\cite{braeckeveldt2024} that curves of Hopf bifurcations converge to a single curve for $\tau\gg\Gamma^{-1}$. We therefore expect no major differences if one were to use an even larger $\tau$. Finally, the only free parameter left is $U$. However, the value of $U$ only determines the minimum value of $A$ for which bistability can be observed. Changing $U$ does not qualitatively affect how the lineshape evolves with the laser-amplitude $A/A_c$.

\clearpage

%\newpage

\renewcommand{\thefigure}{S\arabic{figure}}
\setcounter{figure}{0}

\section{Supplemental Material}

\subsection{Experimental Setup}
\begin{figure*}[!t]
    \centerline{\includegraphics[width=.8\textwidth]{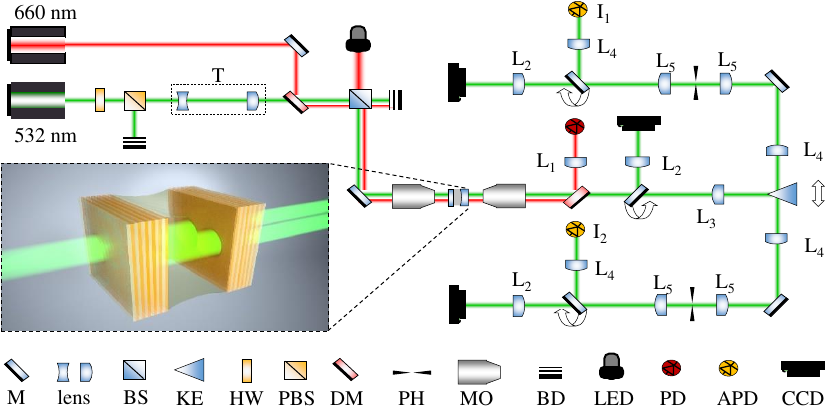}}
	\caption{Schematic of our optical setup. Abbrevations: mirror (M), beam splitter (BS), knife-edge prism (KE), half-wave plate (HW), polarizing beam splitter (PBS), dichroic mirror (DM), pinhole (PH), microscope objective (MO), beam dump (BD), photodetector (PD), CCD camera (CCD). Focal lengths of lenses are $L_1=50$~mm, $L_2=200$~mm, $L_3=1000$~mm, $L_4=75$~mm, and $L_5=100$~mm. Magnification factor of telescope is $T=8\times$. Pinhole size is $150$~$\mu$m. Dichroic mirrors are shortpass mirrors with 605~nm cutoff wavelength. Inset: Schematic of a laser-driven oil-filled optical dimer.}\label{figsup:S1}
\end{figure*}
In this section we outline our experimental setup and measurement procedures. Fig.~\ref{figsup:S1} shows a schematic of our optical setup, comprising a tunable microcavity filled with olive oil. The cavity is made by two mirrors, each comprising a distributed Bragg reflector (DBR) with peak reflectance of 99.8\% at $530$~nm. A dimer is formed by two adjacent concave features (each $5.2~\mu$m diameter and $6~\mu$m radius-of-curvature, and their center-to-center distance $2.7~\mu$m) on one the DBR mirrors~\cite{Trichet15,dufferwiel2015}. In all our experiments, we drive the right cavity by a $532$~nm single-mode laser, and measure the third longitudinal mode. 

We use piezoelectric actuators to align and position the mirrors, and to modulate the planar mirror. The mirror with concave features is mounted on a nanopositioner with six degrees of freedom. The nanopositioner allows us to align the mirrors parallel to each other with microdegree precision, and to position the concave features relative to the pump laser with nanometer precision. We use a high power LED to image in transmission during alignment and positioning. The planar mirror is mounted on a ring piezo with $2.6~\mu$m maximum displacement. In all our experiments the planar mirror is modulated linearly at $20$~Hz over $600$~nm.

Optical excitation (collection) is achieved through a microscope objective with $10\times$ ($20\times$) magnification and a numerical aperture of 0.25 (0.4). Transmission through the two cavities is measured independently on two different avalanche photodetectors by splitting the transmitted beam using a knife-edge prism and isolating the individual concave features using pinholes. A 660~nm diode laser is used to reference the laser-cavity detuning $\Delta_j$ to the cavity dissipation rate $\Gamma$. This is possible because the transmission of our DBR mirrors is high at this wavelength, resulting in a poor cavity for our reference laser. Consequently, the resonances at 660~nm are much broader than for the 532~nm pump laser, and the change in transmission for the 660~nm diode laser is effectively linear over many linewidths of the 532~nm laser. This linear dependence allows us to keep track of $\Delta_j/\Gamma$.

\subsection{Model Validation and Experimental Signatures of Predicted Bifurcations}
\begin{figure*}[!t]
    \centerline{\includegraphics[width=\textwidth]{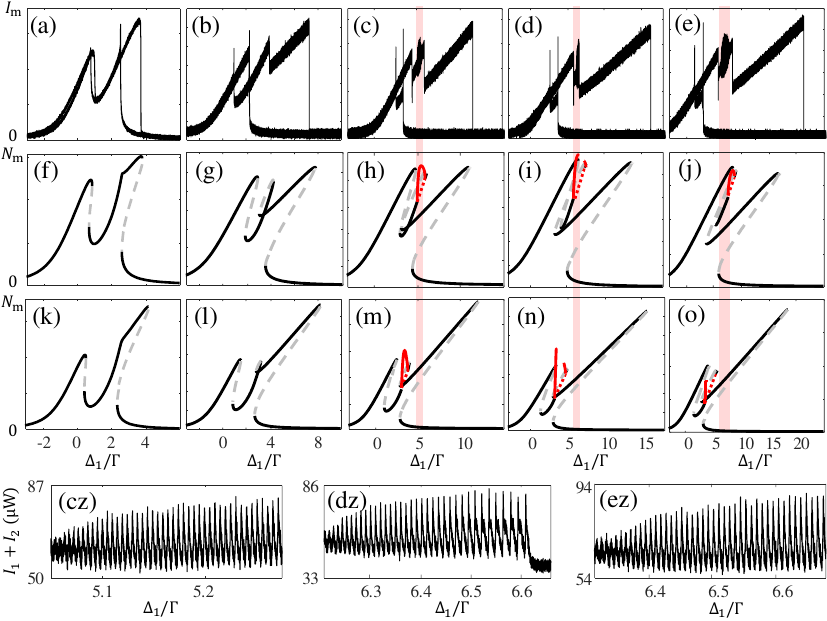}}
    \caption{(a-e) Measured transmitted intensity $I_1+I_2$ versus detuning for increasing pump power. From left to right $A=1.5A_c$, $A=2.7A_c$, $A=3.9A_c$, $A=4.9A_c$ and $A=6.1A_c$, with $A_c$ the critical power for bistability. Shaded red area indicates the detuning range in which oscillations were observed combining all 40 scans. (cz-ez) Zoom into oscillations of (c-e), respectively. (f-j) and (k-o) show steady-state photon number $N_1+N_2$ versus detuning for $\beta=800/\Gamma$ and $\beta=0$, respectively, and driving amplitudes corresponding to (a-e). Solid black: stable; dashed gray: saddle; dotted red: unstable focus; solid red: limit cycle maximum. Model parameters: $\kappa_\mathrm{L}=\Gamma/2$, $U=\Gamma/40$, $\tau=5000\Gamma^{-1}$, $J=\Gamma$, $\delta=-0.25\Gamma$, $\rho=0.95$. Maximum intensities $I_\textrm{m}$ in (a-e) are $37$, $70$, $110$, $110$, and $120~\mu$W, respectively. Maximum photon numbers $N_\textrm{m}$ in (f-j) are $155$, $350$, $500$, $600$, and $800$, respectively. In (k-o) $N_\textrm{m}$ is $170$, $350$, $500$, $700$, and $800$, respectively.}\label{figsup:S2}
\end{figure*}

In this section we demonstrate that the inclusion of a nonlinear loss term in our model is required to reproduce our experimental findings over a large parameter range. Figs.~\ref{figsup:S2}(a-e) show single shot measurements as function of detuning at increasing pump power. The shaded red area indicates the detuning range in which oscillations were observed, combining 40 scans in detuning. Steady-state calculations using $\beta=\Gamma/800$, shown in Figs.~\ref{figsup:S2}(f-j), qualitatively reproduce the experimental observations at all driving powers. On the other hand, setting $\beta=0$ as in Figs.~\ref{figsup:S2}(k-o) does not reproduce our measurements at all driving powers. For $\beta=0$, the detunings at which Hopf bifurcations occur are shifted relative to the measurements by several $\Gamma$. Most notable is the difference between model and experiment at the highest driving power [Figs.~\ref{figsup:S2}(e,o)], for which we observe a wide detuning range with oscillations in experiment, whereas the model with $\beta=0$ predicts only a very small range. Additionally, the model with $\beta=0$ predicts a homoclinic bifurcation at this power, for which we find no evidence experimentally. The model with $\beta=\Gamma/800$, however, predicts the correct bifurcations, and the range in detuning where oscillations occur is much closer to experiments than for $\beta=0$.

\subsection{Limit Cycle Amplitude}
\begin{figure*}[!t]
    \centerline{\includegraphics[width=\textwidth]{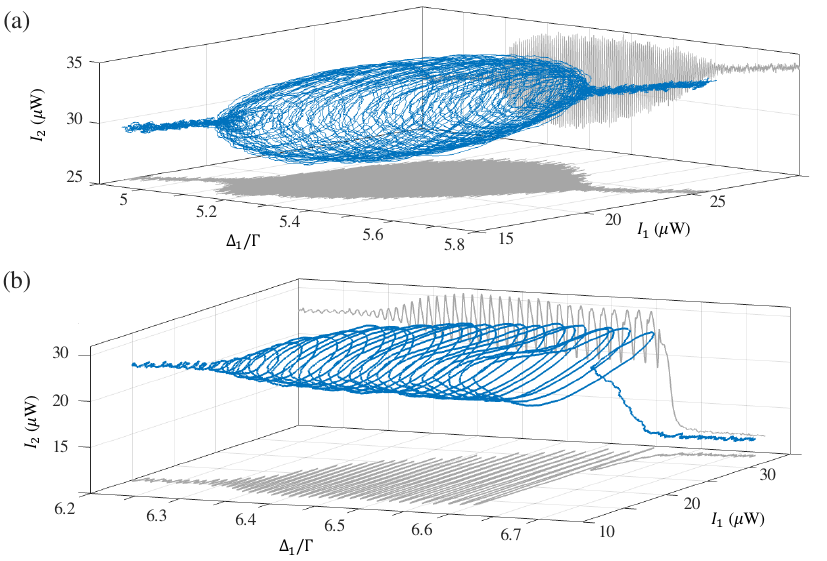}}
    \caption{Average transmitted intensities $I_1$ and $I_2$ as function of $\Delta_1/\Gamma$ for (a) $A=3.9A_c$ and (b) $A=4.9A_c$.}\label{figsup:S_amplitude}
\end{figure*}

In this Section we highlight the difference in behavior of the limit cycle amplitude when approaching a Hopf or homoclinic bifurcation. Figure~\ref{figsup:S_amplitude} shows as 3D plots the measured transmitted intensities $I_1$ and $I_2$ as function of $\Delta_1/\Gamma$, averaged over 40 scans in $\Delta_1/\Gamma$. In Fig.~\ref{figsup:S_amplitude}(a), the region of limit cycles is bound by Hopf bifurcations on both sides. Here, we observe a nonmonotonic behavior of the limit cycle amplitude. On the other hand, in Fig.~\ref{figsup:S_amplitude}(b), the region of limit cycles is bound by a Hopf bifurcation at small detuning and a homoclinic bifurcation at large detuning. In this case, the limit cycle amplitude increases monotonically with $\Delta_1/\Gamma$ until reaching the homoclinic bifurcation, where the oscillations abruptly disappear. After crossing the homoclinic bifurcation the system remains in a stationary state.

\subsection{Precision in Relation to Intracavity Intensity, Oscillation Amplitude and Phase Space Velocity}
\begin{figure}[!t]
    \centerline{\includegraphics[width=\columnwidth]{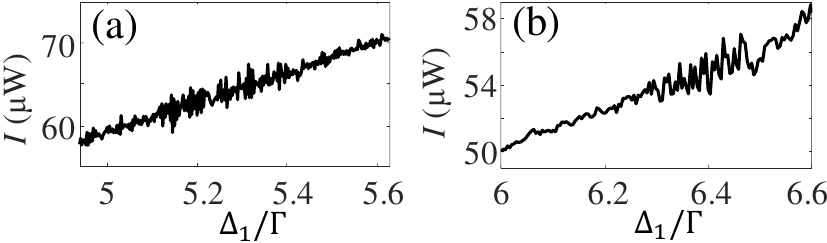}}
    \caption{Average transmitted intensity versus $\Delta_1/\Gamma$ for (a) $A=3.9A_c$ and (b) $A=4.9A_c$.}\label{figsup:S3}
\end{figure}

We formulated several hypotheses to explain our experimental observations. First we thought that the precision enhancement may be due to an increase in the total intracavity intensity $I_1 + I_2$. While the laser power is constant, $I_1 + I_2$ varies with the in-coupling efficiency which depends on  detuning. A large  $I_1 + I_2$  is expected to make fluctuations less relevant at constant noise variance, thereby potentially explaining our observations.  However, Fig.~\ref{figsup:S3} shows that $I_1 + I_2$ behaves qualitatively different from the precision:  $I_1 + I_2$ increases monotonically with the detuning in the self-oscillating regime. This observation rules out our first hypothesis. 

Second, we hypothesized that the precision increases with the oscillation amplitude, which is a measure of the energy in the limit cycle. Large amplitude oscillations could be less influenced by noise, thereby enhancing the precision. However, notice in Figs.~\ref{figsup:S2}(dz) and~\ref{figsup:S_amplitude}(b) that the oscillation amplitude, unlike the precision, increases monotonically with the detuning until reaching the homoclinic bifurcation. This rules out our second hypothesis. 

\begin{figure}[!t]
    \centerline{\includegraphics[width=\columnwidth]{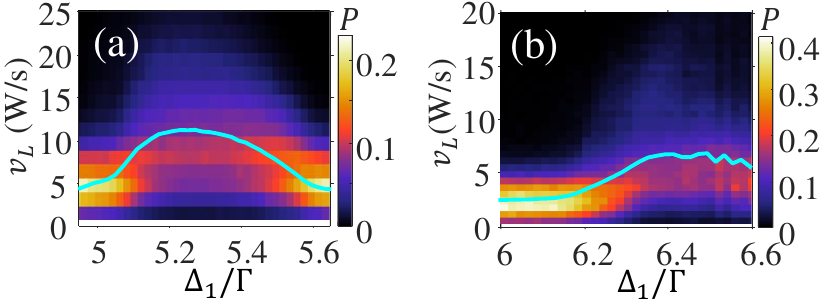}}
    \caption{(e,f) Joint probability distribution of $v_L$ versus $\Delta_1/\Gamma$ for (a) $A=3.9A_c$ and (b) $A=4.9A_c$. Cyan curve shows mean velocity.}\label{figsup:S4}
\end{figure}

Third, we hypothesized that the precision follows the ratio of the amplitude to the period of the limit cycle, which determines its phase space velocity. To this end, we show in Figure~\ref{figsup:S4} the joint probability distribution of the phase space velocity $v_L$ versus $\Delta_1/\Gamma$, with the mean phase space velocity shown in cyan. Although the mean $v_L$ for $A=3.9A_c$ shown in Fig.~\ref{figsup:S4}(a) shows qualitatively the same behavior as the corresponding precision, the behavior of $v_L$ for $A=4.9A_c$ shown in Fig.~\ref{figsup:S4}(b) does not. In particular, Fig.~2(c) of the main text shows a maximum precision around $\Delta_1/\Gamma\sim 5.4$, while $v_L$ has its maximum around  $\Delta_1/\Gamma\sim 5.2$. Similarly, Fig.~2(d) of the main text shows a maximum precision around $\Delta_1/\Gamma\sim 6.3$, while $v_L$ increases until $\Delta_1/\Gamma\sim 6.4$. Additionally, the observed decrease in $v_L$ in Fig.~\ref{figsup:S4}(b) is much smaller than the decrease in the corresponding precision. Therefore, also the observed behavior in $v_L$ is inconsistent with that of the precision.

\end{document}